\documentclass[twocolumn,superscriptaddress,floatfix,preprintnumbers,showpacs,showkeys,pra]{revtex4-1}
\usepackage{graphics,amssymb,amsmath,epsfig,color}
\usepackage{graphicx}

\usepackage{bm}
\usepackage[urlcolor=blue,colorlinks=true,citecolor=blue,linkcolor=blue,pdfstartview={FitH},bookmarks=false]{hyperref}
\usepackage[T1]{fontenc}

\newcommand{\change}[1]{#1}
\begin{document}

\preprint{}
\title{Phase separation of superconducting phases in the Penson--Kolb--Hubbard model}
\author{Konrad Jerzy Kapcia}
\email[corresponding author; e-mail: ]{konrad.kapcia@ifpan.edu.pl}
\affiliation{Division of Physics of Magnetism, Institute of Physics, Polish Academy of Sciences, al. Lotnik\'ow 32/46, PL-02-668 Warszawa, Poland}
\author{Wojciech Robert Czart}
\affiliation{Electron States of Solids Division, Faculty of Physics, Adam Mickiewicz University in Pozna\'n, ul. Umultowska 85, PL-61-614 Pozna\'n, Poland}
\author{Andrzej Ptok}
\email[e-mail: ]{aptok@mmj.pl}
\affiliation{Institute of Nuclear Physics, Polish Academy of Sciences, ul. Radzikowskiego 152, PL-31-342 Krak\'ow, Poland}

\date{May 5, 2015}

\begin{abstract}
In this paper we determine the phase diagrams (for $T=0$ as well as $T>0$) of the Penson-Kolb-Hubbard model for two dimensional square lattice within Hartree-Fock mean-field theory focusing on investigation of superconducting phases and possibility of the occurrence of the phase separation. We obtain that the phase separation, which is a state of coexistence of two different superconducting phases (with $s$-wave and $\eta$-wave symmetries), occurs in define range of the electron concentration. In addition, increasing temperature can change the symmetry of the superconducting order parameter (from $\eta$-wave into $s$-wave). The system considered exhibits also an interesting multicritical behaviour including bicritical points.
\end{abstract}



\maketitle

\section{Introduction}

There has been much interest in high-temperature (unconventional) superconductivity (SC) over more than last two decades.
Moreover, the phase separations phenomenon involving superconducting states (SS) is a very current topic after it has been evidenced in a broad range of intensely investigated materials including iron pnictides, cuprates and organic conductors (discussed below, also see for examples in Refs.~[\onlinecite{ksenofontov.wartmann.11,johnston.10,stewart.11,park.inosov.09,ricci.poccia.11,goko.aczel.09,simonelli.mizokawa.14,xu.tan.08,
udby.andersen.09,savici.fudamoto.02,mohottala.wells.06,pan.oneal.01,kornilov.pudalov.04,colin.salameh.08,taylor.carrington.08,fernandes.schmalian.10}] and references therein).
The phase separation is a  coexistence of two (homogeneous) phases. In such a state coexisting phases form domains, which can differ from each other by, for example, electron concentration or order parameter.
It is worth to note that the phase separations involving superconducting states have been evidenced in a broad range of currently intensely investigated materials including iron pnictides, cuprates and organic conductors.
In particular, there are experimental evidences of phase separation between superconducting and (anti-)ferromagnetic~\cite{ksenofontov.wartmann.11,fernandes.schmalian.10} or magnetic and non-magnetic (superconducting) order~\cite{park.inosov.09,ricci.poccia.11,goko.aczel.09,simonelli.mizokawa.14} in iron-based superconductors.
Moreover,  the coexistence of superconductivity and magnetic order~\cite{xu.tan.08,savici.fudamoto.02,udby.andersen.09} as well as charge-order~\cite{pan.oneal.01} has been reported in cuprates.
Organic compounds also exhibit the superconductor-insulator phase separations as a~result of the external pressure (e.g. quasi-one dimensional (TMTSF)$_2$PF$_6$~\cite{kornilov.pudalov.04} and (TMTSF)$_2$ReO$_4$~\cite{colin.salameh.08}) and fast cooling rate through the glass-like structure transition (e.~g. $\kappa$-(ET)$_2$Cu[N(CN)$_2$]Br~\cite{taylor.carrington.08}).

The Penson-Kolb-Hubbard (PKH) model is one of the conceptually simplest phenomenological models for studying correlations and for description of superconductivity in narrow-band systems with short-range, almost unretarded  pairing~\cite{micnas.rannniger.90,hui.doniach.93,japardze.muller.97,japaridze.kampf.01,robaszkiewicz.bulka.99,czart.robaszkiewicz.01,czart.robaszkiewicz.04,czart.robaszkiewicz.06,czart.robaszkiewicz.15a,czart.robaszkiewicz.15b,dolcini.montorsi.00,robaszkiewicz.czart.01}. The PKH model includes also a nonlocal pairing mechanism (the intersite pair hopping term $J$) that is distinct from the on-site interaction $U$ in the attractive Hubbard model and that is the driving force of pair formation and also of their condensation~\cite{micnas.rannniger.90,robaszkiewicz.bulka.99}. Notice that in the absence of the on-site $U$ term the PKH model reduces to the Penson-Kolb model~\cite{penson.kolb.86,kolb.penson.86}, whereas in the absence of the intersite $J$ term it reduces to the standard Hubbard model~\cite{micnas.rannniger.90,hubbard.63}.

In this paper we determine the phase diagrams (for $T=0$ as well as $T>0$) of the PKH model for two dimensional (2D) square lattice within Hartree-Fock mean-field (HF--MF) theory focusing on a behavior of superconducting phases with changing model parameters and on a possibility of the occurrence of the state with phase separation. We obtain that the phase separation state, in which two different superconducting phases ($s$-wave and $\eta$-wave) can coexists, occurs in a definite range of electron concentration. In addition, increasing temperature can change the symmetry of the superconducting order parameter (from $\eta$-wave into $s$-wave).
The paper is organized as follows. In Section~\ref{sec:model} we present the derivation of HF--MF grand canonical potential and introduce some basic concept of phase separation. Next, Section~\ref{sec:results} is devoted for presentation of the results of numerical computations for the ground state ($T=0$, Section~\ref{sec:groundstate}) and finite temperatures ($T>0$, Section~\ref{sec:finitetemp}). Section~\ref{sec:summary} reports the important conclusions and provides supplementary discussion.

\section{Model and methods}\label{sec:model}

The purpose of the present work is the analysis of phase diagrams, in particular including phase separation, of the extended Hubbard model with pair hopping interaction, i.e. the so-called PKH model, which Hamiltonian in the real space has the following form
\begin{eqnarray}
\label{eq.ham} \mathcal{H} &  =  & \mathcal{H}_{0} + \mathcal{H}_{int}, \\
\label{eq.ham.zero} \mathcal{H}_{0} &=& \sum_{ \langle i,j \rangle \sigma } \left( - t - \mu \delta_{ij} \right) c_{i\sigma}^{\dagger} c_{j\sigma} , \\
\label{eq.ham.int} \mathcal{H}_{int} &=& U \sum_{i} c_{i\uparrow}^{\dagger} c_{i\uparrow} c_{i\downarrow}^{\dagger} c_{i\downarrow} + J \sum_{ \langle i,j \rangle } c_{i\uparrow}^{\dagger} c_{i\downarrow}^{\dagger} c_{j\downarrow} c_{j\uparrow},
\end{eqnarray}
where $c^{\dagger}_{i\sigma}$ ($c_{i\sigma}$) denotes the creation (annihilation) operator of an electron with spin $\sigma$ at site $i$.
$t$ is the single electron hopping integral between nearest-neigbors (NN), $\mu$ is the chemical potential, $U$ is the on-site density-density interaction, and $J$ is the intersite charge-exchange (pair hopping) interaction between NN, respectively.
The electron hopping amplitude ($t$) will be taken as a scale of energy in the system.

The pair-hopping term ($J$) was first proposed in Refs.~[\onlinecite{penson.kolb.86,kolb.penson.86}] by Penson and Kolb in 1986 and it can be derived from a general microscopic tight-binding Hamiltonian, where the Coulomb repulsion may lead to the pair hopping interaction $J =  \langle ii | e^{2} / r | jj \rangle$ \cite{micnas.rannniger.90,czart.robaszkiewicz.01,czart.robaszkiewicz.04,czart.robaszkiewicz.06,czart.robaszkiewicz.15a,czart.robaszkiewicz.15b,robaszkiewicz.bulka.99,kapcia.robaszkiewicz.12,kapcia.robaszkiewicz.13,kapcia.14a,kapcia.14b,kapcia.15a,kapcia.15b,mierzejewski.maska.04}. In such a case $J$ is positive (\emph{repulsive} model $J > 0$, favoring $\eta$-wave SC), but in this case the magnitude of $J$ is very small. However, the  effective attractive form ($J<0$, favoring $s$-wave SC) is also possible (as well as an enhancement of the magnitude of $J>0$) and it can originate from the coupling of electrons with intersite (intermolecular) vibrations via modulation of the hopping integral or from the on-site hybridization term in the general periodic Anderson model (cf. e.g. Ref.~[\onlinecite{micnas.rannniger.90,czart.robaszkiewicz.01,czart.robaszkiewicz.04,czart.robaszkiewicz.06,czart.robaszkiewicz.15a,czart.robaszkiewicz.15b,robaszkiewicz.bulka.99,kapcia.robaszkiewicz.12,kapcia.robaszkiewicz.13,kapcia.14a,kapcia.14b,kapcia.15b,kapcia.15a,mierzejewski.maska.04}] and references therein). It can also be included in  the effective models for Fermi gas in an optical lattice in the strong interaction limit~\cite{rosch.rasch.08}. The role of $J$ interaction in a multiorbital model is of a particular interest because of its presence in the iron pnictides~\cite{ptok.crivelli.15}. It should be stressed that the Hubbard model on bipartite lattice has been rigorously proved to have $\eta$-wave SS states as eigenstates~\cite{yang.1989}. Moreover, $\eta$-wave pairing has been found as a mechanism of superconductivity in a large class of models of strongly correlated electron system (extended Hubbard models)~\cite{boer.korepin.1995}. It has been found that SC originating from the $J>0$ interaction is unique in that it is robust against the orbital (diamagnetic) pair breaking mechanism~\cite{mierzejewski.maska.04}.
The recent studies show that pair-hopping term can also play an important role in nano-space layered structures~\cite{kusakabe.12} and nano-devices~\cite{azema.dare.13}.
Moreover, its effects in various fermionic systems~\cite{matsuura.miyake.12,nishiguchi.kuroki.13,ding.zhang.14,kraus.dalmonte.13,ptok.crivelli.15} have been studied, in particular on the charge-Kondo effect~\cite{matsuura.miyake.12} and Majorana edge states~\cite{kraus.dalmonte.13} as well as in cuprates~\cite{nishiguchi.kuroki.13} and iron-pnictides~\cite{ptok.crivelli.15}. The pair-hopping interactions have been also intensively studied in Bose systems (e.g. Ref.~[\onlinecite{sarker.lovorn.12,zhang.yin.13}]).

In order to analyze superconducting phases we perform the mean-field factorization of interaction Hamiltonian~(\ref{eq.ham.int}):
\begin{eqnarray}
\label{eq.ham_space_mf}
\mathcal{H}^{HF}_{int} &=& U \sum_{i} \left( \Delta_{i}^{\ast} c_{i\downarrow} c_{i\uparrow} + h.c.
\right) - U \sum_{i} | \Delta_{i} |^{2} \\
\nonumber
&+& J \sum_{ \langle i,j \rangle } \left( \Delta^{\ast}_{j} c_{i\downarrow} c_{i\uparrow} + h.c.
\right) - J \sum_{ \langle i,j \rangle } \Delta^{\ast}_{i}  \Delta_{j},
\end{eqnarray}
where we define local superconducting order parameter (SOP) as $\Delta_{i} = \langle c_{i\downarrow}c_{i\uparrow} \rangle$.
In general case, the local SOP is given as $\Delta_{i} = \Delta_{0} \exp ( i {\bm Q} \cdot {\bm r}_{i} )$, where $\Delta_0$ is the amplitude of SOP,   ${\bm Q} = ( 0, 0)$ for $s$-wave SS and ${\bm Q} = ( \pi , \pi)$ for $\eta$-wave SS.
The mean-field Hamiltonian $\mathcal{H}^{HF} = \mathcal{H}_{0} + \mathcal{H}^{HF}_{int}$  in the momentum space takes the form~\cite{ptok.mierzejewski.08,ptok.maska.09}:
\begin{eqnarray}
\label{eq.ham_mom_mf} \mathcal{H}^{HF} &=& \sum_{{\bm k} \sigma } \mathcal{E}_{{\bm k}\sigma} c_{{\bm k}\sigma}^{\dagger} c_{{\bm k}\sigma} \\
\nonumber &+& U_{\bm Q}^{eff} \sum_{\bm k} \left( \Delta_{0}^{\ast} c_{-{\bm k}+{\bm Q} \downarrow} c_{{\bm k}\uparrow} + H.c. \right) - U_{\bm Q}^{eff} N | \Delta_{0} |^{2} ,
\end{eqnarray}
where $\mathcal{E}_{{\bm k}\sigma} = - t \gamma_{\bm k} - \mu$ is a dispersion relation of free electrons (independent of spin $\sigma$ in the absence of external magnetic field, with lattice constants $a=1$ and $b=1$), $\gamma_{\bm k} = 2 \left ( \cos ( k_{x} ) + \cos ( k_{y} ) \right)$ (for 2D square lattice),  $U_{\bm k}^{eff} = U + J \gamma_{\bm k}$ is a effective pairing interaction in the momentum space, and $N=N_{x} \times N_{y}$ is number of sites in the lattice.

Notice that HF-MF Hamiltonians (\ref{eq.ham_space_mf}) and~(\ref{eq.ham_mom_mf}) exhibit the particle-hole symmetry (with respect to half-filling ($n=1$, $\mu=0$), so the phase diagrams will be presented as a function of $|\mu|$ and $|1-n|$~\cite{micnas.rannniger.90,kapcia.robaszkiewicz.12}.
In particular, the phase diagrams are symmetric with respect to $\mu = 0$ or $n=1$.
Let us stress that Hamiltonian (\ref{eq.ham}) also exhibits this symmetry, but at half-filling the corresponding value of chemical potential is $\mu=U/2$, which is exact result for Hamiltonian (\ref{eq.ham}).

\subsection{The Bogoliubov transformation}\label{sec:bogoliubov}

\begin{figure}
\includegraphics[scale=1.2]{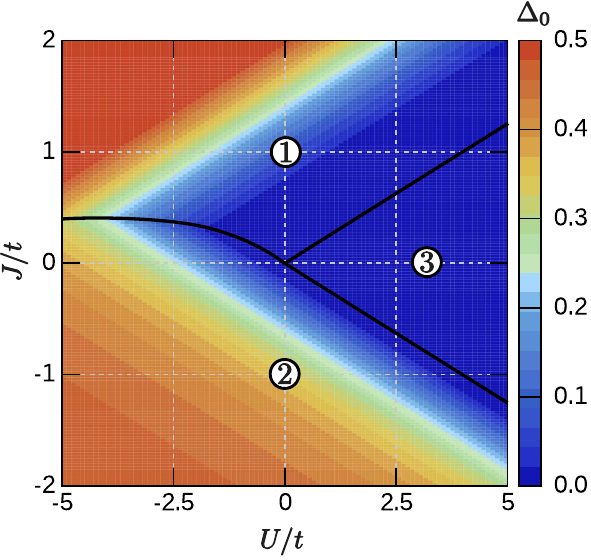}
\caption{Ground state $J/t$ vs. $U/t$ phase diagram for  $n=1$ ($\mu=0$, obtained for $k_{B} T = 10^{-5} t$).
The numbers: $1$, $2$, and $3$ (in circles) denote the regions of occurrence of particular phases: $\eta$-wave SS, $s$-wave SS, and NO, respectively. The transition between regions $1$ and $2$ is discontinuous, whereas the transitions between regions $1$ and $3$ as well as between $2$ and $3$ are continuous.
The colour intensity indicates a value of the amplitude $\Delta_0$ of SOP (compare also with Ref.~[\onlinecite{ptok.kapcia.15}]).
}
\label{fig:fig1}
\end{figure}

In the Nambu notation Hamiltonian~(\ref{eq.ham_mom_mf}) takes the following form:
\begin{eqnarray}
\mathcal{H}^{HF} &=& \sum_{\bm k} \Psi_{\bm k}^{\dagger} H_{\bm k} \Psi_{\bm k} + \mbox{const}
\end{eqnarray}
with
\begin{eqnarray}
\quad H_{\bm k} &=& \left( \begin{array}{cc}
\mathcal{E}_{{\bm k}\uparrow} & U_{\bm Q}^{eff} \Delta_{0} \\
U_{\bm Q}^{eff} \Delta_{0}^{\ast} & -\mathcal{E}_{-{\bm k}+{\bm Q} \downarrow}
\end{array} \right) ,
\end{eqnarray}
where $\Psi_{\bm k}^{\dagger} = ( c_{{\bm k}\uparrow}^{\dagger} , c_{-{\bm k}+{\bm Q} \downarrow} )$ are Nambu's spinors.
Eigenvalues of matrix $H_{\bm k}$ are given by:
\begin{eqnarray}
\lambda_{{\bm k},\pm} = \zeta_{\bm k} \pm \vartheta_{\bm k},
\end{eqnarray}
where plus (minus) sign corresponds to the particle (hole) excitations.
Next, we assume that
\begin{eqnarray}
\vartheta_{\bm k} = \sqrt{ ( \eta_{\bm k} - \mu )^{2} + | U_{\bm Q}^{eff} \Delta_{0} |^{2} } , \\
\zeta_{\bm k} = - t \frac{ \gamma_{\bm k} - \gamma_{-{\bm k}+{\bm Q}} }{2} , \quad \eta_{\bm k} = - t \frac{ \gamma_{\bm k} + \gamma_{-{\bm k}+{\bm Q}} }{2} .
\end{eqnarray}
Matrix $H_{\bm k}$ can be diagonalized using unitary transformation $\mathcal{U}_{\bm k}$, which has the form:
\begin{eqnarray}
\mathcal{U}_{\bm k} &=& \frac{1}{2} \left( \begin{array}{cc}
u_{\bm k} & v_{\bm k} \\
- v_{\bm k} & u_{\bm k}
\end{array} \right),
\end{eqnarray}
where
\begin{eqnarray}
u_{\bm k} &=& \frac{1}{2} \sqrt{ 1 + ( \eta_{\bm k} - \mu ) / \vartheta_{\bm k} } , \\
v_{\bm k} &=& \frac{1}{2} \sqrt{ 1 - ( \eta_{\bm k} - \mu ) / \vartheta_{\bm k} } .
\end{eqnarray}
Then $H_{\bm k} = \mathcal{U}_{\bm k}^{\dagger} \cdot \mbox{diag} ( \lambda_{{\bm k},+} , \lambda_{{\bm k},-} ) \cdot \mathcal{U}_{\bm k}$, and the grand canonical potential takes the form:
\begin{eqnarray}
\nonumber
\Omega &\equiv& -k_{B} T \ln \left\{ {\rm Tr} \left[ \exp(-\beta \mathcal{H}^{HF})\right] \right\} \\
\label{eq.free_ene}
&=& - k_{B} T \sum_{{\bm k},\tau=\pm} \ln \left( 1 + \exp ( - \beta \lambda_{{\bm k}\tau} ) \right) \\
\nonumber &+& \sum_{\bm k} \left( \mathcal{E}_{{\bm k},\downarrow} - U_{{\bm Q}_{SC}}^{eff} | \Delta_{0} |^{2} \right),
\end{eqnarray}
where $\beta = 1 / k_{B} T$ and $T$ is absolute temperature. The electron concentration $n$ in the system is determined from the condition:
\begin{eqnarray}\label{eq.conc}
n & \equiv & \langle n \rangle  \equiv - \frac{1}{N} \frac{d\Omega}{d\mu} \\
\nonumber &=& 1 + \frac{1}{N_{x} N_{y}} \sum_{\bm k} \frac{ \eta_{\bm k} - \mu }{\vartheta_{\bm k}} \left[ f ( \lambda_{{\bm k},+} ) - f ( \lambda_{{\bm k},-} ) \right ]
\end{eqnarray}
where $f ( \omega ) = 1 / ( 1 + \exp ( \beta \omega ) )$ is the Fermi-Dirac distribution.

\begin{figure}[!ht]
\begin{center}
\includegraphics[scale=1.3]{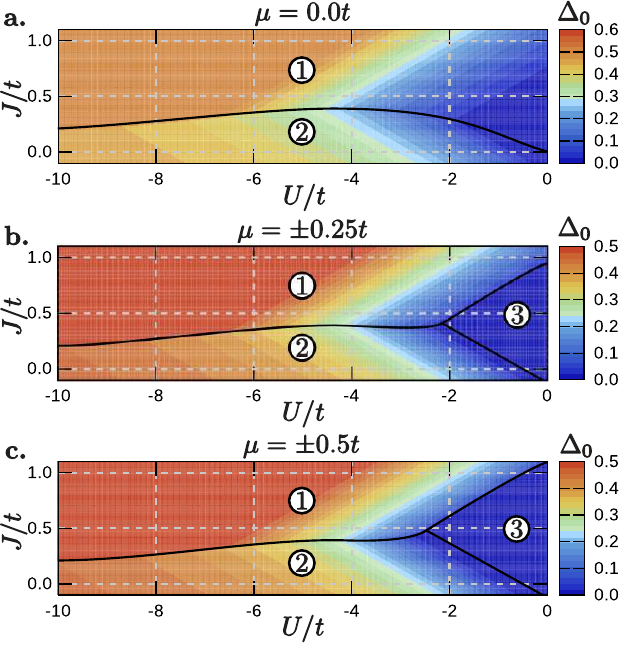}
\end{center}
\caption{
Ground state  $J/t$ vs. $U/t$ phase diagram for  (a) $\mu=0.0t$, (b) $\mu=\pm 0.25t$, (c) $\mu=\pm 0.5t$ ($k_{B} T = 10^{-5} t$).
Denotations as in Fig.~\ref{fig:fig1}.
For $U\rightarrow-\infty$ the discontinuous transition between two SS phases occurs at $J=0$ (independently of $\mu$).
}
\label{fig.dfuj_lim}
\end{figure}

\begin{figure*}
\includegraphics[scale=1]{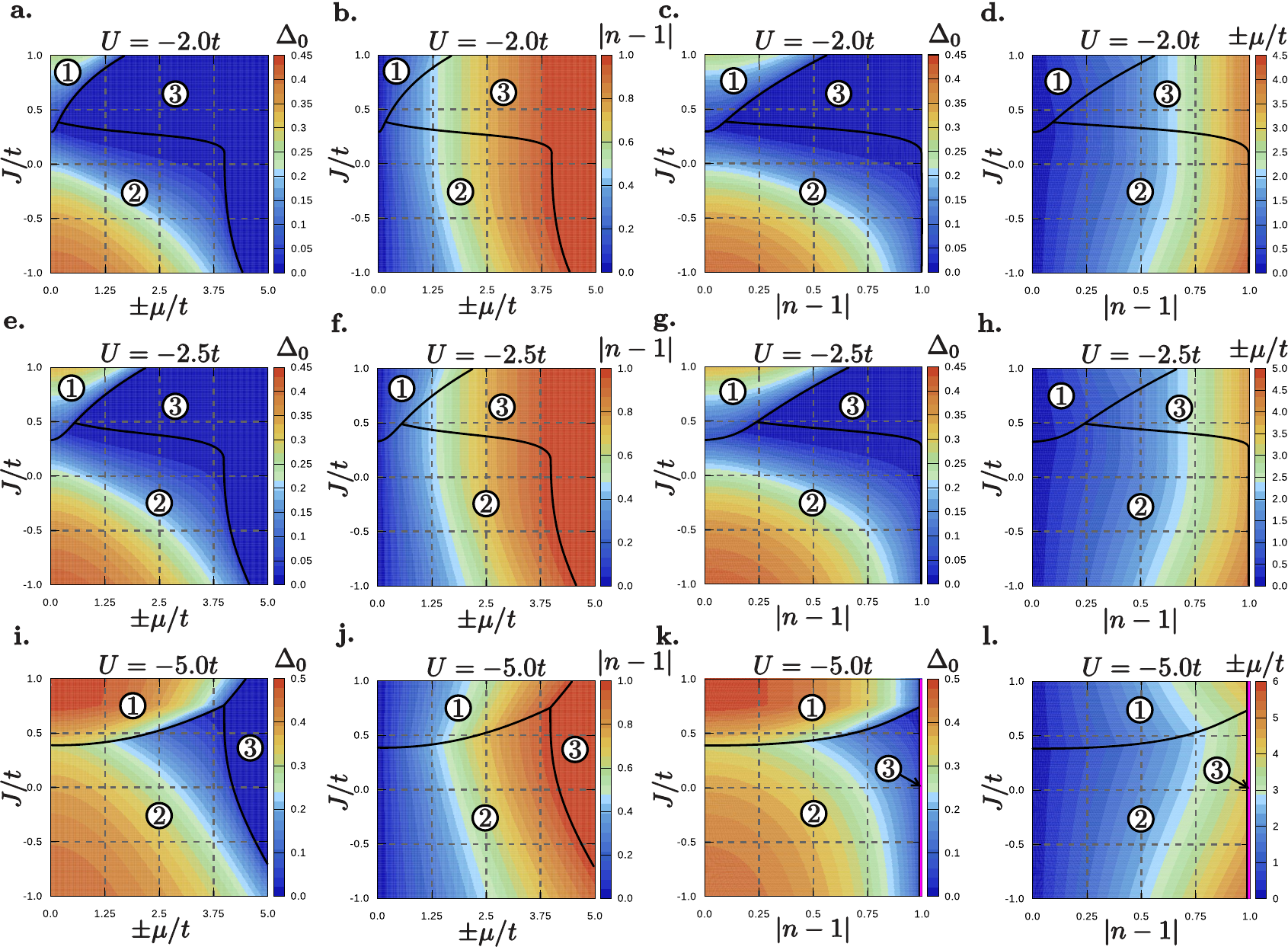}
\caption{Ground state ($ k_{B} T = 10^{-5} t$) phase diagrams $J/t$ vs. $\mu/t$ (first and second column) and $J/t$ vs. $|n-1|$ (third and fourth column) for fixed values of $U/t$ (fixed in all panels in the row), $U/t=-2.0,-2.5,-5.0$, respectively.
The numbers: $1$, $2$, and $3$ (in circles) denote the regions of occurrence of particular phases: $\eta$-wave SS, $s$-wave SS, and NO.
The transition between regions $1$ and $2$ is discontinuous (for fixed $\mu$, only first and second column), whereas the transitions between regions $1$ and $3$ as well as between $2$ and $3$ are continuous.
Between regions $1$ and $2$ there are the regions of the  PS state occurrence (for fixed $n$, only third and fourth column). They are not denoted, because these regions are very narrow  (narrower than thickness of the curves in the figure)
The colour intensity indicates a value of the amplitude $\Delta_0$ of SOP (first and third column), the electron concentration $|1-n|$ (second column) and the chemical potential $|\mu|/t$ (fourth column).
}
\label{fig:fig2}
\end{figure*}

\subsection{The state with phase separation}\label{sec:ps}

In this subsection we would like to introduce the concept of phase separation and introduce the basics of its theory \change{(also cf. e.g. Refs.~[\onlinecite{arrigoni.strinati.91,bak.04,fernandes.schmalian.10,kapcia.robaszkiewicz.12,kapcia.robaszkiewicz.13}]).}
Phase separation (PS) is a state in which two domains with different electron concentration: $n_+$ and $n_-$ exist in the system
(coexistence of two homogeneous phases). The free energies of the PS states are calculated from the expression:
\begin{eqnarray}
\label{row:freeenergyPS}
f_{PS}(n_{+},n_{-}) = m f_{+}(n_{+}) + (1-m) f_{-}(n_{-}),
\end{eqnarray}
where $f_{\pm}(n_{\pm})$ are values of a free energy of two separating phases at $n_{\pm}$ corresponding to the
lowest homogeneous solution for a~given phase ($f=\Omega/N+\mu n$, calculated using~(\ref{eq.free_ene}) and~(\ref{eq.conc})),
$m$ is a fraction of the system with electron concentration $n_{+}$, $1-m$ is a~fraction with electron concentration $n_-$ ($n_{+}>n_{-}$) and
\begin{eqnarray}
\label{eq.PSn}
mn_+ +(1-m)n_-=n,
\end{eqnarray}
where $n$ is fixed.
The minimization of~(\ref{row:freeenergyPS}) with respect to $n_+$ and $n_-$ ($n$ fixed) yields the equality between the chemical potentials in both domains:
\begin{eqnarray}
\label{row:PS1}
\mu_+(n_+)=\mu_-(n_-)
\end{eqnarray}
(chemical equilibrium)
and the following equation (so-called Maxwell's construction):
\begin{eqnarray}
\label{row:PS2}
\mu_+(n_+)=\frac{f_{+}(n_{+})-f_{-}(n_{-})}{n_{+}-n_{-}},
\end{eqnarray}
which is equivalent with equality of grand potentials per site in domains: $\omega_+(\mu_+)=\omega_-(\mu_-)$. It implies that the transitions with a~discontinuous change of $n$ from $n_-$ to $n_+$ in the system considered for fixed $\mu$ can lead to occurrence of the regions of phase separation in the concentration range $n_-<n<n_+$ on the diagrams obtained as a function of $n$. In these regions the homogeneous phases can be metastable as well as unstable, depending on the $n$-dependence of $\mu$. In the PS states the chemical potential $\mu=\mu_+(n_+)=\mu_-(n_-)$ is independent of the electron concentration, i.e. $\partial \mu/\partial n=0$.

\change{%
On the other hand, there is another more intuitive approach. In such an approach the grand canonical potential $\omega = f -\mu n$ is used and chemical potential is independent variable instead of $n$ for free energy $f$, as it was in previous case. In both separating phases chemical potential has the same value. As a consequence we obtain the following simplified  procedure:
at first step we solve the equations for homogeneous phases and next we determined the transition point between both phases (for the condition: $\omega_+(\mu)=\omega_-(\mu)$). Usually, the electron concentrations in these both phases are different ($n_+>n_-$). In such a case, for electron densities between $n_-<n<n_+$ the PS separation state can occur, in which considered phases coexist.}

\change{The PS instability is specific to the short-range nature of the interactions in the model. In the presence of (unscreened)
long-range Coulomb interactions, only a frustrated PS can occur (mesoscale, nanoscale) with the formation of various possible textures  and the large-scale (macroscopic) PS can be prevented \cite{coleman.yukalova.95,yukalov.yukalova.04,yukalov.yukalova.14}.}

\section{Numerical results}\label{sec:results}

All calculations have been performed on graphic processor units using NVIDIA CUDA parallel computing technology, in momentum space on a square lattice grid  $N_x \times N_y = 1000 \times 1000$, using the algorithm described in Ref.~[\onlinecite{januszewski.ptok.14}].

All phase transition boundaries, necessary to construct the complete phase diagram for fixed $\mu$, have been obtained numerically by comparing the grand potential~(\ref{eq.free_ene}) for the solutions found. The transition boundaries for fixed $n$ have been determined by comparing the  free energies $f=\Omega/N+\mu n$ for homogeneous phase (it is calculated by using~(\ref{eq.free_ene}),  the concentration $n$ is determined by~(\ref{eq.conc})) and phase separated states (determined by~(\ref{row:freeenergyPS})). It has been also checked that these results are consistent with the boundaries obtained from the results for fixed $\mu$ (discussed above) by determining the values of electron concentration (equation~(\ref{eq.conc})) on the both sides of transition boundaries derived at fixed $\mu$, which is thermodynamically conjugate to $n$.

\subsection{The ground state ($T=0$)}\label{sec:groundstate}

In this section we discuss the phase diagrams for model~(\ref{eq.ham}) at the ground state ($T=0$).
For $T=0$ on the phase diagrams as a function of the chemical potential $\mu$ the following three homogeneous phases occurs: $s$-wave SS, $\eta$-wave SS and normal (non-ordered, NO) phase.
The diagrams (Figs.~\ref{fig:fig1}--\ref{fig:fig2}, as a function of $\mu$) are nonsymmetric  with respect to $J=0$ and consists of three regions in which phases mentioned above occur.
The $\eta$-wave SS phase  can occur only for $J>0$, whereas  $s$-wave SS phase can be stable for $J<0$ as well as for $J>0$ (in restricted ranges). The transition between both SS phases is discontinuous with a discontinuous change of global SOP defined as $\Delta_{\bm Q} = \tfrac{1}{N} \sum_i \Delta_i \exp(i {\bm Q} {\bm R}_i) $, where ${\bm Q} = (0,0)$ for the $s$-wave SS phase and ${\bm Q} = (\pi,\pi)$ for the $\eta$-wave SS phase ($\Delta_{(0,0)} = \Delta_0$ or $\Delta_{(\pi,\pi)} = \Delta_0$ in the $s$- or $\eta$-wave SS phase, respectively). The transitions between the SS phases ($s$-wave or $\eta$-wave, $\Delta_0 \neq 0$) and the NO phase ($\Delta_0 = 0$) are continuous ones.

The diagram for the half-filling ($n=1$, $\mu=0$) is shown in Fig.~\ref{fig:fig1} (also cf. Ref.~[\onlinecite{ptok.kapcia.15}]).
In the range presented in Fig.~\ref{fig:fig1} the boundary between two SS phases is decreasing function of $U/t$.
Notice that a necessary condition for the SS phases occurrence is $U^{eff}_{i} \leq0$ (or $U^{eff}_{\bm k} \leq 0$), thus the regions of the SS phases must be restricted at least by lines $U \pm 4 J = 0$, which are also the boundaries of the phases occurrence determined by minimization of $\Omega$ for $\mu=0$. However, for a general case of $\mu\neq0$ the SS ($s$- or $\eta$-wave) phases are stable only if $|U^{eff}_i|$ is higher than some critical value and the boundary of stability of the particular phases determined by minimization of $\Omega$ are moved towards lower values of $U/t$ (cf. Fig.~\ref{fig.dfuj_lim}).

In the limit $U/t \rightarrow - \infty$ the discontinuous boundary between both SS phases is located at $J/t = 0$ (independently of $\mu$). In such a limit there is a full symmetry between $s$-wave SS and $\eta$-wave SS phases~\cite{kapcia.robaszkiewicz.12,kapcia.14a,kapcia.robaszkiewicz.13,kapcia.14b,kapcia.15a,kapcia.15b}.
Notice also that in this limit model~(\ref{eq.ham}) is equivalent with the hard-core boson model on the lattice~\cite{micnas.rannniger.90,kapcia.robaszkiewicz.12}.

\begin{figure}
\includegraphics[scale=1]{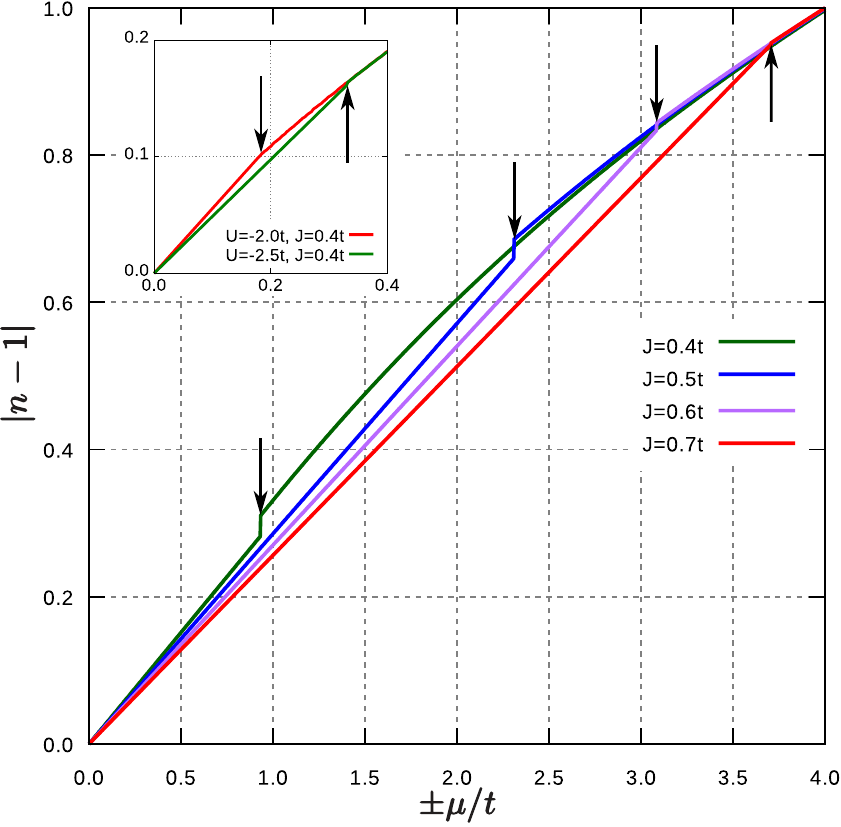}
\caption{The $\mu/t$-dependence of electron concentration $|n-1|$ for $U/t=-0.5$ and various values of $J/t=0.4,0.5,0.6,0.7$ (as labelled) at the ground state ($k_BT=10^{-5}t$). In the inset the $\mu/t$-dependence of electron concentration $|n-1|$ for $J/t=0.4$ and $U/t=-2.0,-2.5$ (as labelled). The discontinuous transitions between two SS phases are indicated by arrows.}
\label{fig:muvsn}
\end{figure}

In Fig.~\ref{fig:fig2} we presents $J/t$ vs. $|\mu|/t$ diagrams (first and second column) as well as $J/t$ vs. $|1-n|$ diagrams (third and fourth column) for the fixed values of $U<0$. The transition between both SS phases for fixed $\mu$ is associated with a discontinuous change of electron concentration $n$ (it is visible especially in Fig.~\ref{fig:muvsn}).
Thus on the phase diagrams as a function of $n$, between the regions of the homogeneous $s$-wave SS and $\eta$-wave SS phases, there are regions of the  PS state occurrence, where both SS phases coexist. In Fig.~\ref{fig:fig2} they are not denoted, because these regions are very narrow  (narrower than thickness of the curves in the figure). In general, the regions of the SS phases occurrence extend with increasing on-site attractive interaction $|U|$, whereas the regions of the NO phases are reduced by increasing $|U|$.

In Fig.~\ref{fig:muvsn} we present the electron concentration $n$ as a function of the chemical potential $\mu$ for fixed model parameters. It is clearly visible that for $U/t=-5$ there is a discontinuity  of $n$ (from $n_-$ to $n_+$) at the transition between two SS phases. The discontinuous transitions between the $\eta$-wave SS and $s$-wave SS phases are indicated by arrows. For $U/t=-2.0,-2.5$ the discontinuity is much smaller, but it still occurs.

\begin{figure*}
\includegraphics[scale=1]{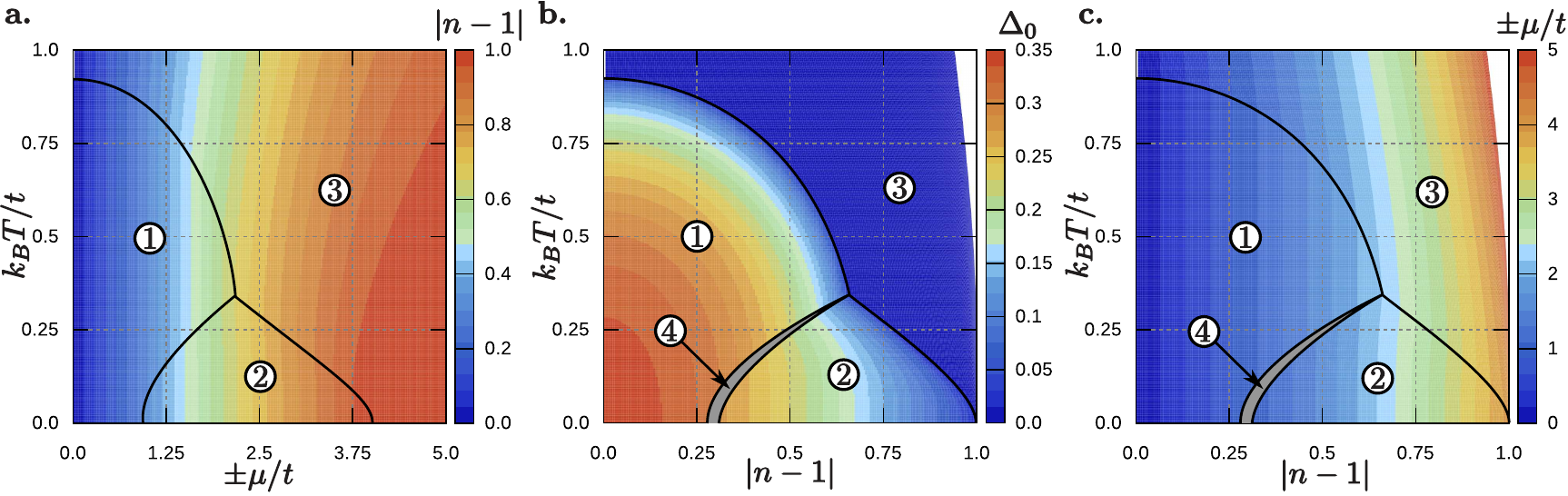}
\caption{(a) $k_BT/t$ vs. $\mu/t$ and  (b), (c) $k_BT/t$ vs. $|n-1|$ phase diagrams for $U/t = -5.0$ and $J/t =0.4$. The colour intensity indicates a value of the electron concentration $|1-n|$ (a), the amplitude $\Delta_0$ of SOP (b), and the chemical potential $\mu/t$ (c).
The numbers: $1$, $2$, $3$, and $4$ (in circles) denote the regions of occurrence of particular phases and states: $\eta$-wave SS, $s$-wave SS, NO, and PS respectively.
The transition between regions $1$ and $2$ is discontinuous (only in panel (a)), whereas the transitions between regions $1$ and $3$ as well as between $2$ and $3$ are continuous.
}
\label{fig:fig5}
\end{figure*}

\subsection{Finite temperatures ($T>0$)}\label{sec:finitetemp}

In this section we discuss the evolution of the phase diagram of the model considered with increasing temperature $T$ and chemical potential $\mu$ (or electron concentration $n$).

As an example, the finite temperature phase diagrams for $U/t=-5.0$ and $J/t=0.4$ are shown in Fig.~\ref{fig:fig5} as a function of $\mu$ and $n$.
On the phase diagrams three homogeneous phases (s-wave SS, $\eta$-wave SS and NO) occur.
The transition between the SS phases and the NO phase are continuous one (second order) and they are decreasing functions of $|\mu|/t$ and $|1-n|$.
The highest transition temperature is for the half-filling ($n=1$, $\mu=0$, the transition from $\eta$-wave SS phase into the NO phase).
The transition from the $s$-wave SS phase to the $\eta$-wave SS phase with increasing temperature is discontinuous (first order) for fixed $\mu$ (Fig.~\ref{fig:fig5}(a)) and its temperature increases with $|\mu|$.
All transition lines (two of second order and one of first order) merge in the bicritical point.
On the phase diagram for fixed $n$ (Figs.~\ref{fig:fig5}(b),(c)), the two SS phases can coexist in the state with (macroscopic) phase separation.
The temperatures of the transitions between the PS state and the homogeneous SS phases increases with $|1-n|$.
In particular, at the ground state ($T=0$) the electron concentrations in the domains are: $n_{-} = 0.2788$ (the $\eta$-wave SS domain) and $n_+ = 0.3125$ (the $s$-wave SS domain).

\begin{figure}
\includegraphics[scale=1.2]{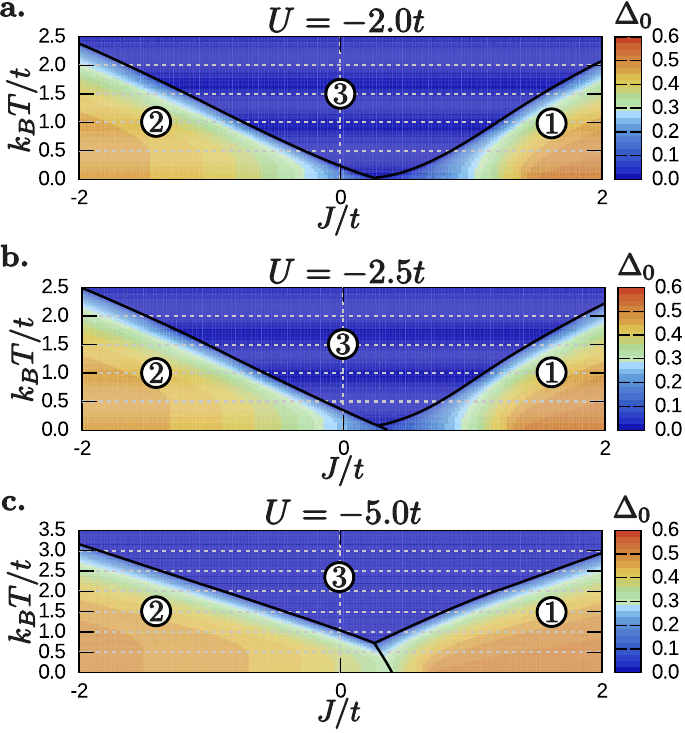}
\caption{The $k_BT/t$ vs. $J/t$ phase diagrams at half-filling ($n=0$, $\mu=0$) for (a) $U/t = -2.0$, (b) $U/t = -3.0$, and (c) $U/t = -5.0$.
Denotations as in Fig.~\ref{fig:fig1}.}
\label{fig:fig6}
\end{figure}

Notice that in the PS state values of $\Delta_0$ (or $\Delta_{\bm Q}$) are undetermined (there are two different order parameters at every domain) and the chemical potential is constant (not dependent on $n$). The different order parameters and electron concentrations in both domains are not dependent on the concentration $n$ of the electrons in the whole system. The general discussion of the PS states properties can be found in e.g. Ref.~[\onlinecite{bak.04,kapcia.robaszkiewicz.12}].

In addition, the phase diagrams for fixed $U/t$ and half-filling ($n=1$, $\mu=0$) are presented in Fig.~\ref{fig:fig6}.
The structure of the diagrams is not dependent on a value of attractive $U$ interaction.
Temperatures of the continuous transition between the $s$-wave SS phase and the NO phase and the discontinuous transition between both SS phase decrease with increasing $J/t$, whereas boundary between the $\eta$-wave SS phase and the NO phase (continuous transition) increase with increasing $J/t$ (for fixed $U/t$).
The regions of the SS phases occurrence extends  with increasing $|U|/t$ and $|J|/t$.
These two interactions induce  superconductivity in the system and also stabilize both SS phases.

Let us stress that  it is possible (for fixed $\mu$ as well for fixed $n$) to change the type of superconductivity occurring in the system  with increasing temperature (the homogeneous $\eta$-wave SS phase can exist in higher temperatures that the homogeneous $s$-wave SS phase, but not contrariwise).

\section{Conclusion and supplementary discussion}\label{sec:summary}

In this paper we studied the superconducting states of the PKH model focusing on the states with phase separation states between two different superconducting phases. We derived phase diagrams of the model and found that two superconducting phases with different symmetry of order parameter can coexist in a state with phase separation. Moreover, the results predict the change of a symmetry of superconducting order (from $\eta$-wave to $s$-wave) with increasing temperature (for fixed $\mu$ as well as fixed $n$).

Notice that one of the results of this paper that the temperature can change the symmetry of superconductivity pairing is consistent with other works~\cite{aperis.kotetes.11,thomale.platt.11a,thomale.platt.11b,maiti.korshunov.11,fernandes.millis.13,livanas.aperis.15}. One of the real materials, which exhibit a pronounced fragility of the gap symmetry, are the iron-based SCs (for review see e.g. Ref.~[\onlinecite{johnston.10,stewart.11,livanas.aperis.15}]).
In fact, recent works have demonstrated the possibility of gap symmetry transitions~\cite{aperis.kotetes.11,thomale.platt.11a,thomale.platt.11b,maiti.korshunov.11,fernandes.millis.13,livanas.aperis.15}, independently of the pairing mechanism. The small-$\vec{q}$ electron-phonon interaction~\cite{aperis.kotetes.11} and the spin-fluctuations scenario~\cite{thomale.platt.11a,thomale.platt.11b,maiti.korshunov.11,fernandes.millis.13} are both compatible, but gap symmetry transitions constitute a characteristic feature of the first mechanism, which leads to a loss of rigidity of the gap function in momentum space (momentum decoupling)~\cite{oppeneer.varelogiannis.03}.

One should also notice that it was also derived that increasing magnetic field can change symmetry of SOP from $s$-wave (or $d$-wave) into $\eta$-wave~\cite{ptok.maska.09}.

Let us stress that it has been reported that superconductivity can coexist with charge-ordered and magnetically ordered phases in states with phase separations~\cite{kapcia.12,kapcia.13,kapcia.15b,kapcia.15a,kapcia.robaszkiewicz.12}.
In present work we do not consider the charge and magnetic orderings. In the PKH model, they can both occur for $U > 0$ (charge ordering can be also present for $U < 0$)~\cite{micnas.rannniger.90,robaszkiewicz.bulka.99,japardze.muller.97,japaridze.kampf.01,japaridze.sarkar.02,japaridze.sarkar.02}.

Although the existence of $\eta$-wave SS has not been confirmed experimentally, the recent theoretical results (e.g. Ref.~[\onlinecite{mierzejewski.maska.04,ptok.kapcia.15,czart.robaszkiewicz.15a,czart.robaszkiewicz.15b,czart.robaszkiewicz.15b}]) give the explicit suggestions how it can be distinguished from the $s$-wave SS. The main issue in the differentiate these two phases explicitly is that only absolute value of the order parameter (energy gap) can be measured experimentally (e.g. by STM spectroscopy) and only the behaviour of thermodynamical properties~\cite{mierzejewski.maska.04,czart.robaszkiewicz.15a,czart.robaszkiewicz.15b,czart.robaszkiewicz.15b} or the influence of impurities on superconducting properties~\cite{ptok.kapcia.15} can give information about the SS symmetry.


\begin{acknowledgments}
The authors thank Stanis\l{}aw Robaszkiewicz for very fruitful discussions and comments.
K.J.K. is supported by National Science Centre (NCN, Poland) -- the grant No. DEC-2013/08/T/ST3/00012 in years 2013--2015.
\end{acknowledgments}


\end{document}